\documentstyle[11pt]{article}

\newcommand {\q} {BR\,1202$-$0725}
\newcommand {\etal} {{et al.}\/~}

\newcommand {\apgt} {\ {\raise-.5ex\hbox{$\buildrel>\over\sim$}}\, }
\newcommand {\aplt} {\ {\raise-.5ex\hbox{$\buildrel<\over\sim$}}\, }

\newcommand {\ks} {{\,km\,s$^{-1}$}}
\newcommand {\I} {~{\protect\sc i}}
\newcommand {\II} {~{\protect\sc ii}}
\newcommand {\III} {~{\protect\sc iii}}
\newcommand {\IV} {~{\protect\sc iv}}
\newcommand {\V} {~{\protect\sc v}}
\newcommand {\IL} {~{\protect\sc i}\,$\lambda$}
\newcommand {\IIL} {~{\protect\sc ii}\,$\lambda$}

\newcommand {\IVL} {~{\protect\sc iv}\,$\lambda$}

\renewcommand{\date}{\today}
\newcommand{\runningtitle}{ABUNDANCES ABOVE Z=4.}
\vspace{1cm}

\makeatletter
\def\section{\@startsection {section}{1}{\z@}{3.5ex plus 1ex minus
    .2ex}{2.3ex plus .2ex}{\Large\bf}}
\def\subsection{\@startsection{subsection}{2}{\z@}{3.25ex plus 1ex minus
   .2ex}{1.5ex plus .2ex}{\large\bf}}
\def\subsubsection{\@startsection{subsubsection}{3}{\z@}{3.25ex plus
1ex minus .2ex}{1.5ex plus .2ex}{\normalsize\bf}}
\makeatother

\begin{document}
\vspace*{3cm} \begin{center}
{\large \bf High Resolution Observations of the QSO BR1202-0725:\\
          Deuterium and Ionic Abundances at Redshifts Above z=4.}\\
\vspace*{2cm}

\begin {center}

E.\ Joseph Wampler\footnote[1]{European Southern Observatory,
Karl-Schwarzschild-Str. 2, D-85748 Garching bei M\"{u}nchen, Germany. Present
address: National Astronomical Observatory, Osawa, Mitaka, Tokyo 181, Japan.},
G.\ M.\ Williger\footnote[2]{Max-Planck-Institut f\"{u}r Astronomie,
K\"{o}nigstuhl 17, D-69117, Heidelberg, Germany.},
 J.\ A.\ Baldwin\footnote[3]{Cerro Tololo Interamerican Observatory, Casilla
603, La Serena, Chile, operated by the Association of Universities for Research
in Astronomy Inc. (AURA) under cooperative agreement with the National Science
Foundation.},
 R.\ F.\ Carswell\footnote[4]{Institute of Astronomy, Madingley Road,
Cambridge, CB3 0HA, UK.},\\
 C.\ Hazard\footnote[5]{Department of Physics and Astronomy, University of
Pittsburgh, 100 Allen Hall, Pittsburgh, PA, 15260, USA.} and
 R.\ G.\ McMahon\footnotemark[4]
\end {center}

\vspace*{3.5cm}

Submitted to Astronomy and Astrophysics, Main Journal, Extragalactic Astronomy
section. \\
Thesaurus code numbers: 11.17.1, 11.17.4 (BR\,1202$-$0725), 12.03.3\\
Running head: Wampler et al., Abundances Above $z = 4$.
\bigskip

\bigskip
\hspace{1cm}Received \_\_\_\_\_\_\_\_\_\_\_\_\_\_\_\_\_\_\_\_\_\_\_\_\_;
accepted \_\_\_\_\_\_\_\_\_\_\_\_\_\_\_\_\_\_\_\_\_\_\_\_\_ \\

\bigskip
\bigskip

   \begin{bf}
     Address for correspondence:\\
   \end{bf}
\smallskip
 J.Baldwin, Cerro Tololo Interamerican Observatory, Casilla 603, La Serena,
Chile.

\end{center}

\newpage

\noindent
{ \bf ABSTRACT. } We present results from 12\ks\ resolution echelle
spectroscopy of the bright $z=4.694$ QSO \q~.  A preliminary analysis
shows that high metallicity narrow line absorption clouds are present
up to the redshift of the quasar.  A damped Ly$\alpha$ system with an
HI column density of $3.1 \times 10^{20}$ cm$^{-2}$ at $z=4.383$ has an
[O/H] ratio that is about 0.01 solar, while another absorption system
with an HI column density of $5.0 \times 10^{16}$ cm$^{-2}$ at
$z=4.672$, may have an O/H ratio that is twice solar. An upper limit,
or the possible detection of deuterium in this absorption cloud gives a
D/H ratio of about 1.5\,10$^{-4}$.  Because this cloud is metal rich at
least some of the cloud gas has been processed after the beginning of
the Universe.


\baselineskip 19pt

\noindent

\section{Introduction}
Element abundances at high redshift are currently of extreme interest because
it is believed that at high enough redshift the element ratios will
disclose the nature of the first generation of stars and their initial
enrichment of the primordial gas clouds. As very efficient spectrographs
that can achieve resolutions in excess of 10\ks are now in regular
use at a number of observatories, it has become practical to undertake
detailed studies of the metallicity of absorption line clouds along
lines of sight to distant quasars. Previous studies (see, for example,
Petitjean \etal 1994)
have shown that some quasar
absorption-line clouds are surprisingly metal rich, even at
very high redshift. However, Matteucci and Padovani (1993) have pointed
out that that in its initial starburst, a galaxy can raise the metallicity of
primeval clouds to nearly the solar value. They suggest that the
metal enrichment of the primordial clouds might take
only 5--8\,$10^8$ years. Thus very high redshifts might be needed in order
to reach look--back times that are sufficiently large that metal rich
clouds are no longer found.

Also important is the abundance of deuterium at high redshift. As it is
believed that deuterium is produced during the formation of the Universe
and subsequently destroyed by the cycling of the initial gas clouds
through stellar processes, the D/H ratio at very high redshift might
be more representative of the initial deuterium abundance than that
found in interstellar clouds in the Solar neighborhood. The new
spectrographs now, for the first time, permit an investigation of the D/H
ratio at high redshift. See Songaila \etal (1994) and Carswell
\etal (1994) for previous determinations of Deuterium abundances at
$z=3.3201$.

The brightest known quasar with a redshift above $z=4.5$ is one found
during the APM optical survey for QSOs with $z>4$ (Irwin, McMahon and
Hazard, 1991). It is \q, which has an emission line redshift of $z=4.694$
and magnitude R=18.7.
The coordinates of this object are given in McMahon et al. (1994). Here
we report on a study of absorption line systems in this object, and
examine three systems with redshifts that range from $z=4.38$ to $z=4.69$.
These are the highest redshift absorption line clouds for which
abundance studies have been reported. A previous study of this quasar
(Giallongo \etal\ (1994)) used somewhat lower resolution to investigate
the Gunn-Peterson break at the onset of the Ly$\alpha$ forest.\\

\section{Observations}
\q~was observed using the ESO NTT telescope and EMMI in December 1992,
January 1993 and April 1994. It was also observed with the CTIO 4-meter
telescope and Cassegrain echelle spectrograph at Tololo in April, 1993. For a
description of the EMMI, see D'Odorico (1990).
The CTIO setup was similar to that in
Williger {\em et al.} (1994).
EMMI can give a spectral resolution of
about 8\ks, while the CTIO spectrograph gives a spectral resolution of
about 12\ks. Integration times for the individual observations ranged
from 2 hours to 3 hours. All these spectra, which have a total integration
time of about 25 hours, were first transformed to vacuum wavelengths at
the Solar system barycenter and then weighted by the quality of the
data and combined to form the final spectrum which is analyzed here.

A Th-A lamp was used to determine the wavelength calibration of
the echelle spectrum. Once a global wavelength fit to the two dimensional
spectrum was found, that fit was used as a trial starting solution to
identify calibration lines in the individual orders. Only unblended,
moderately strong Th-A lines were used in the fits. Typically 6-10 good lines
were found for each order and the resulting RMS deviation of these lines
from the third order polynomial fit was about 3-5 milliangstroms. EMMI
has very good wavelength stability. Wavelength drift caused by flexure or
temperature changes during a two hour exposure is expected to be less than
about 2\ks. The largest single uncertainty in the wavelength calibration
is the uncertainty in the distribution of starlight in the 1.5 arcsec slit.
If our uncertainties combine quadraticly,
the velocity integrity of the spectrum
should be good to about 2\ks, and we believe that
the zero point of the wavelength scale
has a similar accuracy. We have relied on the accuracy of this calibration
over a wide wavelength range in the analysis of the spectrum. This has
provided a substantial constraint on possible models of the spectrum.

By first shifting our spectra to barycenter wavelengths, we have widened
and distorted the lines in our spectrum that are due to the Earth's
atmosphere. It was difficult to correct the individual spectra for the
atmospheric lines because we do not have observations of suitable
standard stars taken at the
same time and airmass as our quasar
observations. We have made a crude correction of the atmospheric b-band
using data from an earlier spectrum of Q\,0059$-$2735 (see Wampler \etal,
1995). Q\,0059$-$2735
had been observed in a very similar fashion to
\q. Because Q\,0059$-$2735 is brighter than \q, the signal to noise ratio
for Q\,0059$-$2735 is higher than that of
\q. Furthermore, in the spectrum of Q\,0059$-$2735 there are no strong lines
in atmospheric b-band spectral region.  Therefore, the spectrum
of Q\,0059$-$2735 can be used to estimate the atmospheric absorption
line contribution to the spectrum of \q. This correction is not perfect,
but it is sufficient to our purposes.

Fig. 1 gives an overview of the spectrum of \q\
(see also the spectrum, taken
with lower resolution and published by Giallongo \etal 1994).
Here the data have
been rebinned to a resolution of 2\AA. The spectrum has been normalized to
a median filtered spectrum of Hiltner 600 (Stone 1977). This procedure
grossly distorted the intensities near the wide atmospheric A-band,
where the filter did not bridge the
band, but did not significantly affect the relative intensities at other
wavelengths where the absorption lines are narrower and the filter
was able to smooth across the lines. In Fig. 1 the plotted intensities
are in wavelength units relative to H 600.

The redshift of BR1202$-$0725 has been determined by Storrie-Lombardi
et al. (1995, ApJ submitted) to be 4.694$\pm$0.010 based on the blue
edge of the Ly$\alpha$ emission line.  This is consistent with the peak
of the Ly$\alpha$ emission line from our high resolution spectrum.
O\IL1304, Si\IVL1400~and C\IVL1549 are the other
detected emission lines (see Fig.\,1). As BR1202$-$0725 is too
heavily absorbed to use the relatively weak metal emission lines
to get an accurate redshift, Storrie-Lombardi \etal used absorption
features to determine the QSO redshift.  This yields redshifts of 4.676
and 4.681 for OI(1304) and CIV(1549) respectively.  BR1202$-$0725 has
been detected at mm and sub-mm wavelengths (McMahon et al., 1994, Isaak
et al., 1994). This emission indicates that there may be large
quantities of dust within the host galaxy of BR1202$-$0725.  We note
that in quasar spectra the high ionization metal lines are often
blueshifted with respect to Ly$\alpha$ (Gaskell 1982; Tytler and Fan
1992).

In this paper we investigate 3 metal absorption line systems in
the spectrum of \q. These are the damped Ly$\alpha$ system at
$z=4.383$, and two systems near the emission-line redshift
of \q. These latter two have redshifts of $z=4.672$, and $z=4.687$,
respectively. There are a number of other metal-line absorption
systems in the spectrum of \q which we have not studied in detail.
The more prominent of these are listed in Table 1. In this paper
we concentrate on the three systems with redshifts of
$z=$4.383, 4.672, and 4.687.\\

\section{Abundance Determinations}
Column densities of the observed lines were found by modeling the
spectrum using a MIDAS procedure built around a MIDAS application
program called CLOUD which was written by M. Pierre and is a
derivative of the program ATLAS (Pettini {\em et al.} 1983).
A description of the procedure used to model
the lines is given by Wampler \etal (1995). The oscillator strengths used
were taken from the compilation by Morton (1991).
This procedure does not fit
the spectrum, but many lines and components can be incorporated into the
model and the resulting curve can be compared with the observed spectrum.
For instance, in modeling the Lyman series in \q, we have used a 50-level
H atom to simultaneously model the lower Lyman lines
and the transition
to the continuum level. Only a few of the higher Lyman lines are
unblended and clearly seen in the data, but the ones that are present,
together with the limits on the optical depth of the Lyman continuum
and the profiles of Ly$\alpha$~and Ly$\beta$ are enough to constrain
both the hydrogen $b$-values and the hydrogen column density.

It is necessary to determine the local continuum level in order to
successfully model the absorption line profiles. To the red of the
Ly$\alpha$~forest, the continuum was
easily determined by running the spectrum through a
20\AA~wide median filter. But in the Ly$\alpha$~forest region, Fig. 1
shows that the local continuum is increasingly depressed as one moves to
the violet. This effect is probably caused by the blending of hundreds of
Ly$\alpha$~forest lines. In the Ly$\alpha$~forest region we estimated
the local continuum level by assuming that the highest locations in a
100\AA~wide sliding band were near the continuum level. Points, selected
this way, were tied together with a spline fit and the resulting smooth
curve was taken to be the local continuum. The echelle spectrum of \q~was
then divided by this continuum and the resulting normalized spectrum was
used to compare with the model absorption spectrum. This procedure
compensates for the continuum depression caused by overlap of many weak
Ly$\alpha$~forest lines, but, because at a particular wavelength the local
continuum may be higher than our depressed continuum, the strengths of
unsaturated lines may be underestimated.

For wavelengths that are to the violet of the \q~Ly$\alpha$ emission peak,
the density of Ly$\alpha$ absorption lines is so great that most
observed absorption features are blends of more than one line. This
makes modeling those metal resonance lines that lie in the
Ly$\alpha$ forest difficult or impossible. Also, the redshift of
\q~shifts the metal lines that lie to the red of Ly$\alpha$ into a
wavelength region where there are many absorption lines belonging to
the Earth's atmosphere. This limits our ability to determine the
metallicity and ionization conditions for many of the metal-line systems
found in the spectrum of \q. In this paper
we report on three high redshift metal-line systems for which it is
possible to find enough metal lines that are sufficiently free of
blends with other lines to permit useful constraints to be placed on
the absorption clouds. In Table 2 we list the transitions and
approximate wavelengths of those transitions in the three systems that we
modeled. Other lines, which can only be used to place upper bounds on
the ionic column densities, are not listed in Table 2. In Table 3 we
give the hydrogen and ionic column densities, cloud velocities and
Doppler
line widths ($b$-values) for these three metal line systems.
For several species our upper limits to the
ionic column densities are low enough to set useful constraints
to possible ionization models. We have also listed these ions in Table 3.

The following subsections contain brief descriptions of these three high
redshift absorption line systems. In reading the following discussion,
note that our models use fairly high $b$-values for several of the
metal line components. Sometimes these components are clearly not
saturated, yet they seem very broad. Probably these are blends of
several sub-components that higher S/N and/or higher spectral resolution
could resolve. However, as Jenkins (1986) has pointed out, ensembles
of moderately saturated lines can be analyzed as if there were only
a single component. Such an analysis gives reasonably correct results
so long as the characteristics of the component clouds are not markedly
irregular. With our present set of data we feel that it is best to
limit the number of component clouds (and the degrees of freedom) in
our model while recognizing that the actual situation may
be more complex than our simplified absorption models.\\

\subsection{System A, $z = 4.383$}
The highest column density H\I~system in \q~is a low ionization system
that contains several components clustered around a redshift of
$z=4.383$. We are designating it ``System A''. Lines of H\I, O\I,
Si\II~and C\II~are strong, while Si\IV~and C\IV~are weak or absent at our
detection limit (column densities of about 5\,10$^{12}$  cm$^{-2}$ for typical
oscillator strengths). Fig. 2 shows our model fits to hydrogen Ly$\alpha$
and several of the metal lines in System A.
Si\III~ and N\V~are in the
extremely rich Ly$\alpha$ forest region of \q. Given that fact that
Si\IV~and C\IV~are so weak, we are doubtful that we have detected N\V.
Rather, it is likely that the feature at the expected location of N\V~is
an unrelated line in the Ly$\alpha$ forest. Similarly, Si\III~may
be blended with Ly$\alpha$ forest lines and may not be as strong
as we suggest, although the presence of Si\III~would not be as
surprising as that of N\V. We have not with certainty detected {\em any}
high ionization lines in System A.

The velocity components listed in Table 3
are those used to model the line profiles,
but the true velocity
structure is somewhat uncertain. We have used model fits with as few as four
absorption components, and find that, unless there are some unresolved
narrow saturated structures, the abundance estimates do not depend
strongly on the number of assumed clouds.

The Si\II~column densities are based on the measurement of two lines:
Si\IIL1304 and Si\IIL1527. These two lines limit our freedom in choosing
individual cloud $b$-values and velocities. Given these constraints, we can
attempt to model the saturated O\I~and C\II~lines. The column densities
listed in Table 3 are the {\em minimum} required to model the profiles.
Saturation could possibly increase the column densities by a factor of 5
if the relative line strengths in O\I~and C\II~differed from those of Si\II.

Only the total column density of H\I~in System A
is known, not the column density of the individual clouds. The
column density of H\I~in System A was determined by requiring that the
damping wings of H\I~Ly$\alpha$ not lie below the intensity of the
``continuum'' region between strong lines in the Ly$\alpha$ forest. In
this region of the spectrum our S/N ratio is about 20:1. The apparent
``noise'' seen in Fig. 1 is actually due mostly to many weak absorption
lines. Comparing the total column density of O\I~with that of H\I, we
have estimated the abundance ratio of O/H to be about 8.3\,10$^{-6}$, or
about 0.01 solar. However, if the O\I~line is more saturated
than we have thought, or if undetected lines in the Ly$\alpha$ forest
have resulted in our overestimating the strength of the damping
wings of the System A Ly$\alpha$, then the metal abundance could be higher
than our estimate.

The CII/OI and SiII/OI ratios are close to the solar abundance ratios
for these elements. This suggests a photoionization model which has
carbon and silicon totally dominated by the singly ionized components,
so a very low ionization parameter gives the best overall fit. We used the
photoionization code {\sc Cloudy} (Ferland 1995) to investigate
this possibility. For a
background flux of $6 \times 10^{-21}$ erg cm$^{-2}$ Hz$^{-1}$ s$^{-1}$
sr$^{-1}$ the hydrogen density is 300 cm$^{-3}$ or more if we treat the
complex as a uniform density system. However, the absence of CII fine
structure absorption suggests this is an overestimate by at least two
orders of magnitude, and the presence of SiIV in two components is not
consistent with this model.  If we take the view that the SiII
measurement is the most reliable (it is based on two lines), and
assume a hydrogen density of 0.3 cm$^{-3}$, then using the
background-photoionized model we find SiII/HI $\sim
1.6\times$Si/H. SiIV is present
at about 1/10 the SiII column density, as for the $z=4.38290$ and
4.38382 components. If the OI column density is correct, then O/Si
would be $\sim 1.6\times$ solar.

We do not expect the models to be accurate to much better than a factor
of two in differential abundances, and the column densities themselves
are not very precise, so we do not regard such discrepancies as at all
serious. In any case a small amount of dust depletion of silicon (and
carbon) by an amount of order the ionization correction would readily
explain the relative column densities seen for a heavy element
abundance $\sim 0.01$ solar.

Since Al is more strongly depleted than Si in the galactic interstellar
medium, a detection here with Al/Si similar to the solar value would
argue against significant dust depletion of either species.
Unfortunately Al\IIL1670 falls in a region which is badly affected by
night sky, but it is possible to determine upper limits to the
Al\II~column density of $\sim 3.0\, 10^{12}$
for the components at $z=4.38382$,
4.38444 and 4.38509. Since these are upper limits, we have no handle on
the dust depletion in this system. The centers of the C\IIL1334 and O\IL1302
lines are saturated, so their column densities are uncertain and
could be higher than the values quoted in the table. With dust
depletion of Al and Si, the overall heavy element abundances could be as
high as 1/25 solar.\\

\subsection{System B, $z = 4.672$}
System B appears to be dominated by absorption from a single ionized
cloud at a redshift of $z=4.67231$. In contrast to System A, the
ionization conditions of System B favor high ionization clouds.
Si\IV~is relatively strong compared to H\I. C\III, Si\III~and C\IV~also
seem to be present, but C\III~and Si\III~lie in the Ly$\alpha$ forest
of \q~ and are blended with lines from other redshift systems.  C\IV~is
in a spectral region that has many strong atmospheric absorption lines.
It is therefore difficult to measure the strengths of these lines
precisely, but there are strong features at the position of the
strongest component of the CIV doublet and at the wavelengths of the
C\III~and Si\III~ lines. Identification of the absorption features with
these metal lines gives reasonable column densities.

The column density of H\I~in System B is quite low. This is often the
case in highly ionized clouds, but there is a feature at the expected
position of redshifted O\I. There is no strong atmospheric line at this
position, nor have we found an alternate identification for the line.
If our identification of this feature with oxygen is correct, then the
O/H ratio for System B is $\approx1/500$, or twice solar. However, at the
present time this identification must remain in doubt, since we have not
found any evidence that C\II~is present. In a high ionization system
the column density of C\II~often exceeds that of O\I, because
C\II~can exist in a H\II~region, while O\I~cannot.
The upper limit to the
column density of C\II~in System B is 10$^{13}$ cm$^{-2}$, 0.1 that of O\I.
In contrast, we note that the C\II/O\I~ratio in System A, a low
ionization system, is 0.3, only slightly higher than the solar ratio.
This might be expected in a low ionization system where the extent of
the H\II~zone is small compared to the extent of the H\I~zone. Such an
explanation is not likely in the case of System B, as it is a high
ionization system. Nevertheless, if our O\I~identification is correct,
it would appear that not only does System B have nearly solar
oxygen abundances, but also the C/O abundance ratio is low compared
to the Sun.
We note that oxygen is an $\alpha$-chain element that builds up in
abundance very rapidly in the early universe, while carbon is an
element that is co-produced with iron and builds up more slowly
(Timmes et al., 1995). Thus, an overabundance of oxygen in a high
redshift system might signal the extreme youth of the metal-line cloud.
Petitjean \etal (1994) have commented that oxygen seems to
be overabundant relative to carbon in the $z=2.034$ system of
PKS\,0424-131. They also find that absorption systems with high metal
content are usually found to have redshifts close to that of the background
quasar. They argue that such systems are physically associated with the
quasar. If our identification of O\I~in System B is correct, this
may be another example of a high metallicity system with anomalous
abundances having a redshift close to that of the background quasar.
However, if the observed absorption feature is not due to O\I, the
metal abundance of System B might be quite low.\\

\subsection{System C, $z = 4.687$}
System C, a cloud complex near redshift $z=4.687$, is, like system B, a
high ionization system with strong C\IV~compared to the low ionization
lines. C\III~may also be present in this system as there is a pair
of lines at the correct wavelength to give a reasonable identification
with C\III.
Fig. 4 shows that the $^5$D\,--\,$^7$P Fe\III\ $\lambda$1214.56 transition
may be present with measurable
strength. However, this line is blended with a strong
line in the atmospheric b-band. In spite of our attempt to correct
the spectrum of \q~for absorption by the b-band, this blend causes an
uncertainty in the column density of Fe\III. Also, the probability of
finding a hydrogen line at the expected position of Fe\III~is
substantial, though the observed feature is somewhat too narrow to be
identified as a hydrogen line. As we have found no evidence for O\I~in
the System C clouds, and as we have no trustworthy ionization model,
we cannot determine the metal abundance in this system. Our absorption
model for System C uses two components: a strong, blue component with
a fairly low $b$-value, and a red component that is weaker and wider than the
blue component. More components could have been used,
but our data in the C\IV~region, where the components are best
seen, are too noisy to justify the increased complexity.

\subsection{The D/H ratio}

Because the high ionization clouds in System B are dominated by a single
absorption component, it is well situated for investigating the
D/H ratio at high redshift. Fig. 4 shows the Ly$\alpha$ and
Ly$\beta$ lines and the higher order Lyman lines for both System B
and System C.

The hydrogen column densities of Systems B and C are constrained by the
observational data in three ways: (1) by the width and shape of
Ly$\alpha$ and Ly$\beta$; (2) by the point at which the higher Lyman
lines stop being saturated; and (3) by the residual continuum level
below the Lyman break.

The widths and shapes of the first two Lyman lines are
relatively unaffected by blends with other lines, as can be seen in
Figure 4.
The point at which the Lyman lines stop being saturated can be judged
from Figure 5, which shows an expanded view of the spectral region
where the transition from the Lyman lines to the Lyman continuum in the
two systems occurs. While the high order Lyman lines fall in a spectral
region that is highly confused by blends with other features, there
are, nevertheless, numerous sections of the \q~spectrum where the
System B and C Lyman lines leave a clear imprint. The Lyman-15 transition
is designated by the numeral 15 directly under the appropriate system.
Since in our
model, System C has two components, we have placed the letter
designating lines in this system between the two components. Even
though the location of the appropriate continuum is uncertain, the fact
that there is little residual intensity until we reach absorption lines
that are high in the Lyman series means that the Lyman lines are
saturated until $\sim$Ly15. Because the $b$-value required to fit the
lower Lyman lines decreases as the assumed column density
increases, the model lines in the higher Lyman series strengthen
and narrow as the column density increases.

Figs. 1 and 5 demonstrate that there is significant residual continuum
in the spectrum of \q~below the Lyman limits of these two absorption
systems. In fact, there is substantial residual continuum shortward of the
System B limit at 5180\AA~to about 5000\AA~where the flux is cut off
by a different system. However, because the exact location
of the local continuum is uncertain, the depression of this continuum
by the Lyman continuum absorptions of systems B and C is probably
a less secure method of determining the hydrogen
column density than noting that we see saturated Lyman lines to very
high n-values. The presence of these saturated higher Lyman lines sets
a lower limit to the hydrogen column density of the clouds, while the
residual continuum in the spectrum of \q~sets an upper limit to this
column density.
For the hydrogen column densities found for systems B and C,
the depth of the continuum absorption should be about 60\% of
the continuum level
in the Lyman series. That this amount of absorption is about correct can
be best seen in the overview given in Fig. 1 (the spectrum of \q\ in this
region is not as noisy as it may appear; there are a very large number of
blended absorption lines in this spectral region and they contribute
substantially to the irregularity in the spectral flux).

Constrained by the observed Lyman continuum flux, the strengths of the
higher Lyman lines, and the profiles of Ly-$\alpha$~and Ly-$\beta$,
and taking into account the first 50 Lyman lines, we iteratively
modeled the hydrogen spectra of Components B and C. Because the Lyman
continuum of System C overlaps the higher order Lyman lines and continuum
of System B, we first determined the bounds on the system C hydrogen
column density and then with the System C continuum in place, we
adjusted the continuum level for System B. We were thus
able to model System B in the presence
of absorption by System C. Once the hydrogen column densities had been
established for the two systems, we adjusted the component $b$-values
(one for System B and two for System C) in order to provide acceptable
models for the lower Lyman lines. The $b$-values required to model the
observed profile are very sensitive to the assumed column densities, so
it is not possible to estimate the column density using only the lower,
saturated, Lyman lines. In \q~we are fortunate that the spectrum gives us
enough constraints to determine the required parameters. We believe that
the model shown in this figure and listed in Table 3 is conservative.
Models with twice the hydrogen column density would not fit as well,
but might be acceptable. Decreasing the H\I~column density by factor of
two would require a large number of chance coincidences of unrelated
absorption lines with the positions of the Lyman series lines.
We feel that such a large number of coincidences would be very
unlikely.

The observed Ly$\alpha$ lines of both System B and System C show
small dips at the expected location of D\I. These positions
are designated in Fig. 4. Because there is a possibility that weak,
satellite H\I~features cause the depressions, we can only determine
upper limits for the D/H ratios in these clouds. The position of
D$_{Ly\alpha}$ in System B is free of strong atmospheric b-band
absorption lines, but the D$_{Ly\alpha}$ line in System C is blended
with an atmospheric b-band line.
Of course, in System C the only D\I~line that we could see is the one
due to the blue (and strongest) component of the two lines in our model.
The D\I~line belonging to the red component is lost in the absorption
due to the blue H\I~component. It is unfortunate that not only is the
D$_{Ly\alpha}$ in System C blended, but the D$_{Ly\beta}$ line in System C
is also confused with other absorption features. We find that
D/H \aplt1.5\,10$^{-4}$ for System B and D/H \aplt10$^{-3}$ for System C,
for the reasons just given, the latter value is extremely uncertain.
In calculating the D/H ratios, we have taken our best estimate for the
hydrogen column densities. If we had taken the upper limit, we could
have decreased the D/H ratios by a factor of about two.
Without better data it is not possible to make a stronger
statement. However, we note that the D/H ratio derived for System B is
similar to that (\,\aplt2\,10$^{-4}$)
found by Songaila \etal (1994) and
Carswell \etal (1994) for the quasar Q\,0014+813.

The upper limit that we find for the D/H ratio in System B (\aplt10$^{-4}$)
is very close to the primordial ratio predicted by the calculations
for the first creation of the light elements during the formation of the
Universe. The initial D/H ratio depends on the baryon-to-photon ratio
$\eta$~and decreases as $\eta$~increases. As the primordial gas is
processed by stars, the D/H ratio can only decrease with time as it is
believed that there are no efficient processes to increase the
deuterium abundance after the big bang. For a discussion of the
decrease of D/H with time
see (Vangioni-Flam \etal, 1994, and Steigman and Tosi, 1992).
The value of $\eta$~can be determined from the primordial $^4$He mass
fraction (Y$_p$). For Y$_p\approx0.23\pm0.01$ (Olive \etal,
1991; Pagel \etal, 1992) $2.8\,10^{-10}\leq\eta\leq4.2\,10^{-10}$
(Vangioni-Flam \etal, 1994). Calculations show that for
$\eta=4$\,10$^{-10}$, (D/H)$_p$=5\,10$^{-5}$. this increases
to (D/H)$_p$=8\,10$^{-5}$ for $\eta=3$\,10$^{-10}$.

Our systems B and C are at very large redshifts, they are thus
``pre-solar''. However, we do not know the fractional gas content of each
cloud that is primordial. For instance, they might even be supernova
remnants, in which case the D/H ratio
should be very small. However, as
both clouds show a weak feature near the expected position of D\I~we may
have actually detected D\I. Clearly these two systems merit further
study, to improve both the H\I~and D\I~column densities. An increase in
the S/N ratio at Ly$\beta$ might lead to a detection of D Ly$\beta$
in System B. This would improve
the accuracy of the column
density measurement of D\I~and might
provide an independent estimate of cloud $b$-values for the light atoms.

\section{Conclusions}
We find that even at redshifts in excess of 4, it is possible to find
intervening quasar absorption systems with strong metal lines. We
identify three such systems, A ($z = 4.383$), B ($z=4.672$) and C
($z=4.687$).
Of the three systems studied here, System A
has the most reliable abundance determination. We estimate that for it,
the O/H abundance is at least $\sim$0.01 times solar.
System B ($z = 4.672$) may be even more metal rich if our identification of
an isolated weak line with oxygen is correct. The O/H abundance in System B
may be $\sim$2 times solar (although this would seem to require a very low
C/O abundance ratio).

High metallicity at large redshift requires a short delay time ($\tau_{delay}$)
between the start of the universe and the beginning of galaxy formation. Timmes
et al.
(1995) have used published quasar abundances to argue that the data are
compatible with $\tau_{delay}$
of the order of 3 Gyr for universes with $\Lambda=0$
and $\Omega$~between 0.2 and 1. From our data $\tau_{delay}$ must be less than
1 Gyr if their assumptions concerning the Universe are correct. In fact,
the metallicity we find in the three absorption systems that we have analyzed
in
\q\ is not qualitatively different from the systems found at
much lower redshift.

Deuterium may have been found in System B. If so, the abundance is
similar to that found earlier in the $z=3.32$ system of Q\,0014+813 by
Songaila \etal (1994) and Carswell \etal (1994).

\section{Acknowledgments}
We would like to thank the ESO La Silla night assistants, in particular
Jorge Merandez and Manuel Bahamondes for their help. We are grateful to
Pascal Ballester, Klaus Banse and Michele Peron for willing help in modifying
the MIDAS programs so that they ran faster and were well adapted to our
echelle data reductions. This paper was completed while EJW was spending
5 months at Beijing Observatory. He is grateful to Professor Jian-shung
Chen and Observatory Director Qi-bin Li
for providing a hospitable climate
at the Beijing Observatory for the writing of this paper.

\newpage

 \begin{center}
 \begin{bf}

     REFERENCES\\
 \end{bf}
 \end{center}
\begin{verse}

Carswell R.F., Rauch M., Weymann R.J., Cooke A.J., Webb J.K., 1994,
MNRAS 268, L1

D'Odorico S., 1990, {\em The Messenger} 61, 51

Ferland G.J., 1995, {\em Hazy, a Brief Introduction to Cloudy},
University of Kentucky Physics Department Internal Report

Gaskell C. M., 1982, ApJ 263, 79

Giallongo E., D'Odorico S., Fontana A., et al., 1994, ApJ 425, L1

Irwin M.J., McMahon R.G., Hazard C., 1991, in: Crampton D. (ed.), ASP
Conf. Ser. Vol. 21, The Space Distribution of Quasars. Astron. Soc. Pac.,
San Francisco, p. 117.

Jenkins E.B., 1986, ApJ 304, 739

Matteucci F., Padovani P., 1993, ApJ 419, 485

McMahon R.G., Omont A., Bergeron J., Kreysa E.,  Haslam C.G.T., 1994,
MNRAS 267, L9

Morton D.C., 1991, ApJS 77, 119

Olive K.A., Steigman G., Walker T.P., 1991, ApJ 380, L1

Pagel B.E.J., Simonson E.A., Terlevich R.J., Edmunds, M.G.,
1992, MNRAS 255, 325

Petitjean P., Rauch M., Carswell R.F., 1994, A\&A 291, 29

Pettini M., Hunstead R.W., Murdoch H.S., Blades, J.C., 1983, ApJ
273, 436

Songaila A., Cowie L.L., Hogan C.J., Rugers, M., 1994, Nature 368,
599

Steigman G., Tosi M., 1992, ApJ 401, 150

Stone R.P.S., 1977, ApJ 218, 767

Storrie-Lombardi L., McMahon R.G., Irwin M.J.,  Hazard C., 1995, ApJ, submitted

Timmes F.X., Lauroesch J.T., Truran J.W. 1995, ApJ 451, 468

Tytler D., Fan X.-M., 1992, ApJS 79, 1

Vangioni-Flam E., Olive K.A., Prantzos N., 1994, ApJ 427, 618

Wampler E.J, Chuga\v{i} N.N., Petitjean, P., 1995, ApJ 443, 586

Williger G.M., Baldwin J.A., Carswell R.F., et al., 1994, ApJ 428, 574

\end{verse}

\pagebreak

\vspace*{3cm}

\begin{center}
{\sc Table 1}~Metal-Line Systems in \q

\bigskip

\begin{tabular}{lclcc} \hline\hline
 & & & & \\
$\lambda_{obs,\,vac}$ & W$_{obs}$ & ION & $\lambda_{lab}$ & $z$ \\
 & & & & \\
\hline
 & & & & \\
7152.6 & 2.4 & Fe\II~ & 2586 & 1.7546 \\
7162.6 & 3.3 & Fe\II~ & 2600 & 1.7546 \\
7702.7 & 5.5 & Mg\II~ & 2796 & 1.7546 \\
7722.8 & 5.4 & Mg\II~ & 2803 & 1.7546 \\
8428.4 & 1.8 & Mg\II~ & 2796 & 2.0141 \\
8450.1 & 0.6 & Mg\II~ & 2803 & 2.0141 \\
8068.5 & 2.9 & Fe\II~ & 2344 & 2.4420 \\
8172.2 & 1.6 & Fe\II~ & 2374 & 2.4420 \\
8201.4 & 3.7 & Fe\II~ & 2382 & 2.4420 \\
8900.4 & 2.4 & Fe\II~ & 2586 & 2.4420 \\
8948.5 & 4.5 & Fe\II~ & 2600 & 2.4420 \\
7070.8 & 1.1 & Si\IV~ & 1393 & 4.0734 \\
7116.7 & 0.6 & Si\IV~ & 1402 & 4.0734 \\
7854.9$^1$ & 3.0 & C\IV~  & 1548 & 4.0734 \\
7867.7 & 1.8 & C\IV~  & 1550 & 4.0734 \\
8384$^2$ & 1.0 & C\IV~  & 1548 & 4.4126 \\
8394$^2$ & 0.5 & C\IV~  & 1550 & 4.4126 \\
8486$^2$ & 3.0 & C\IV~  & 1548 & 4.4811 \\
8499$^2$ & 1.8 & C\IV~  & 1550 & 4.4811 \\
 & & & & \\
\hline\hline
\end{tabular}
\end{center}
\small
{\hspace{3.5cm}$^1$Noise spike in line.}

{\hspace{3cm}$^2$At least two velocity components,
structure not determined.}
\normalsize

\pagebreak

\vspace*{2cm}

\begin{center}
{\sc Table 2.}~List of Modeled Line Complexes\\

\bigskip
\begin{tabular}{lcccccc} \hline\hline
\multicolumn{7}{c}{\sc System A $z=4.383$}\\
\hline
 & & & & & & \\
$\lambda_{obs,\,vac}$ & 6544 & 7184 & 7010 & 7022 & 8219 & 6494 \\
W$_{\lambda}^{obs}($\,\AA ) & 65~~ & 3.76 & 3.97 & 1.21 & 2.20 & 3.33 \\
Ident & Ly$\alpha$ & C\II & O\I & Si\II & Si\II & SiIII \\
 & & & & & & \\
\hline
\multicolumn{7}{c}{\sc System B $z=4.672$}\\
\hline
 & & & & & & \\
$\lambda_{obs,\,vac}$ & 6896 & 5818 & 7386 & 7906 & 8782 & \\
W$_{\lambda}^{obs}($\,\AA ) & 3.11 & 2.15 & 0.33 & 0.47 & 0.64 & \\
Ident & Ly$\alpha$ & Ly$\beta$ & O\I & Si\IV & C\IV & \\
 & & & & & & \\
\hline
\multicolumn{7}{c}{\sc System C $z=4.687$}\\
\hline
 & & & & & & \\
$\lambda_{obs,\,vac}$ & 6914 & 5833 & 6908 & 7927 & 8806 & \\
W$_{\lambda}^{obs}($\,\AA ) & 6.38 & 4.49 & 0.4: & 0.83 & 3.44 & \\
Ident & Ly$\alpha$ & Ly$\beta$ & Fe\III & Si\IV & C\IV & \\
 & & & & & & \\
\hline\hline
\end{tabular}
\end{center}

\pagebreak

\vspace*{2cm}

\begin{center}
{\sc Table 3.}~Cloud Column Densities\\

\bigskip

\begin{tabular}{clccccccc} \hline\hline
\multicolumn{1}{c}{\sc Sys.} & \multicolumn{1}{c}{$Z$} &
\multicolumn{1}{c}{$b$(H\I)} & \multicolumn{1}{c}{H\I$^1$} &
\multicolumn{1}{c}{D\I} &
\multicolumn{1}{c}{$b$(metals)} & \multicolumn{1}{c}{C\II} &
\multicolumn{1}{c}{C\III} & \multicolumn{1}{c}{C\IV} \\
\hline
& & & & & & & \\
A & 4.38133 & & & & 10 & 3.3E13 & & \aplt1E13 \\
A & 4.38192 & & & & 10 & 3.2E13 & & \aplt1E13 \\
A & 4.38250 & & & & 10 & 5.6E13 & & \aplt1E13 \\
A & 4.38290 & & 3.1E20 & & 12 & 3.6E14 & & \aplt1E13 \\
A & 4.38382 & & & & 30 & 3.8E14 & & \aplt1E13 \\
A & 4.38444 & & & & 20 & 8.0E13 & & \aplt1E13 \\
A & 4.38509 & & & & 12 & 8.0E13 & & \aplt1E13 \\
& & & & & & & \\
B & 4.67231 & 22 & 5.0E16 & \aplt8.0E12 & 15 & & \aplt2.0E14 & 4.0E13: \\
& & & & & & & \\
C & 4.68664 & 25 & 1.4E16 & \aplt2.5E13 & 20 & & 2.0E14: & 2.0E14 \\
C & 4.68831 & 45 & 7.2E15 & & 30 & & 5.0E13: & 1.5E14 \\
& & & & & & & \\
\hline\hline
\multicolumn{1}{c}{\sc Sys.} & \multicolumn{1}{c}{$Z$} &
\multicolumn{1}{c}{$b$(metals)} & \multicolumn{1}{c}{N\V} &
\multicolumn{1}{c}{O\I} &
\multicolumn{1}{c}{Si\II} & \multicolumn{1}{c}{Si\III} &
\multicolumn{1}{c}{Si\IV} & \multicolumn{1}{c}{Fe\III} \\
\hline
& & & & & & & \\
A & 4.38133 & 10 & & 8.8E13 & \aplt4.0E12 & \aplt1.5E12 & & \\
A & 4.38192 & 10 & & 8.0E13 & \aplt4.0E12 & \aplt2.0E12 & & \\
A & 4.38250 & 10 & & 1.4E14 & 7.0E12 & \aplt3.0E14 & & \\
A & 4.38290 & 12 & \aplt8.0E13 & 9.0E14 & 4.5E13 & \aplt6.0E13 &
\aplt1.0E13 & \\
A & 4.38382 & 30 & \aplt8.0E13 & 9.6E14 & 4.8E13 & \aplt2.0E13 &
\aplt1.0E13 & \\
A & 4.38444 & 20 & & 2.0E14 & 1.0E13 & \aplt1.0E12 & & \\
A & 4.38509 & 12 & & 2.0E14 & 1.0E13 & \aplt5.0E12 & & \\
& & & & & & & \\
B & 4.67231 & 15 & & 1.0E14 & & \aplt8.0E12 & 1.2E13 &  \\
& & & & & & & \\
C & 4.68664 & 20 & & & & \aplt3.0E14 & 1.5E13 & \aplt1E16\\
C & 4.68831 & 30 & & & & \aplt2.0E12 & 5.0E12: & \aplt2E15 \\
& & & & & & & \\
\hline\hline
\end{tabular}
\end{center}
\small{$^1$Only the total column density of H\I~in System A is
known, not its distribution among the individual clouds.}
\normalsize

\pagebreak

\vspace*{2cm}

 \begin{center}
   \begin{bf}
     FIGURE CAPTIONS\\
   \end{bf}
 \end{center}

\begin{description}
\item[Fig. 1.\,---] Overview of the spectrum of \q~rebinned to 2\AA~per bin.
The positions of several strong emission lines are marked.
9 pixels with values below -1 have been assigned the value -1, and 1
pixel with a value greater that 18 has been assigned the value 18.
The flux is relative to that of Hiltner 600.

\item[Fig. 2.\,---] The damped Ly$\alpha$~line of the low ionization
system A, together with several of the metal lines. Because there
is severe blending with Lyman lines in other redshift systems,
the N\I\ ,
N\V~and Si\III~features are considered only to be upper limits
to the true column densities.

\item[Fig. 3.\,---] Selected metal absorption lines in systems B and C.

\item[Fig. 4.\,---] The Lyman series of H\I~for systems B and C. The
maximum column density of system C is set by the requirement there not
be too much absorption in the interval between 5180\,\AA~and 5200\,\AA.
The maximum column density of system B is set by a similar requirement
for the Lyman continuum region below 5180\,\AA. The numeral ``15'' indicates
the location of the Lyman-15 transition.

\item[Fig. 5.\,---] Details of the higher Lyman lines for
Systems B and C. The numeral ``15'' indicates the location of the
Lyman-15 transition.
\end{description}

\begin{figure}
\vspace*{14cm}
\includegraphics{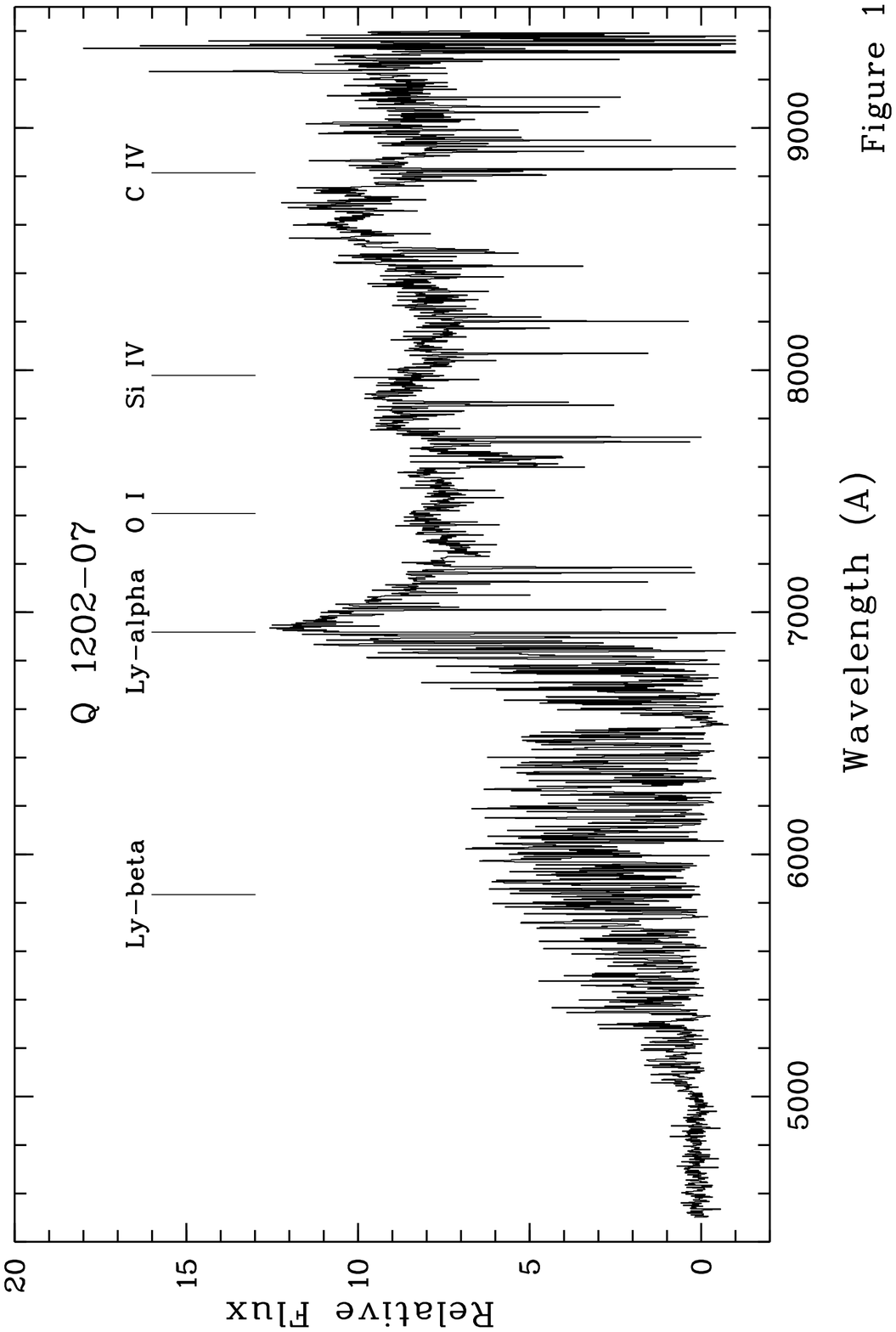}
\end{figure}

\begin{figure}
\vspace*{14cm}
\includegraphics{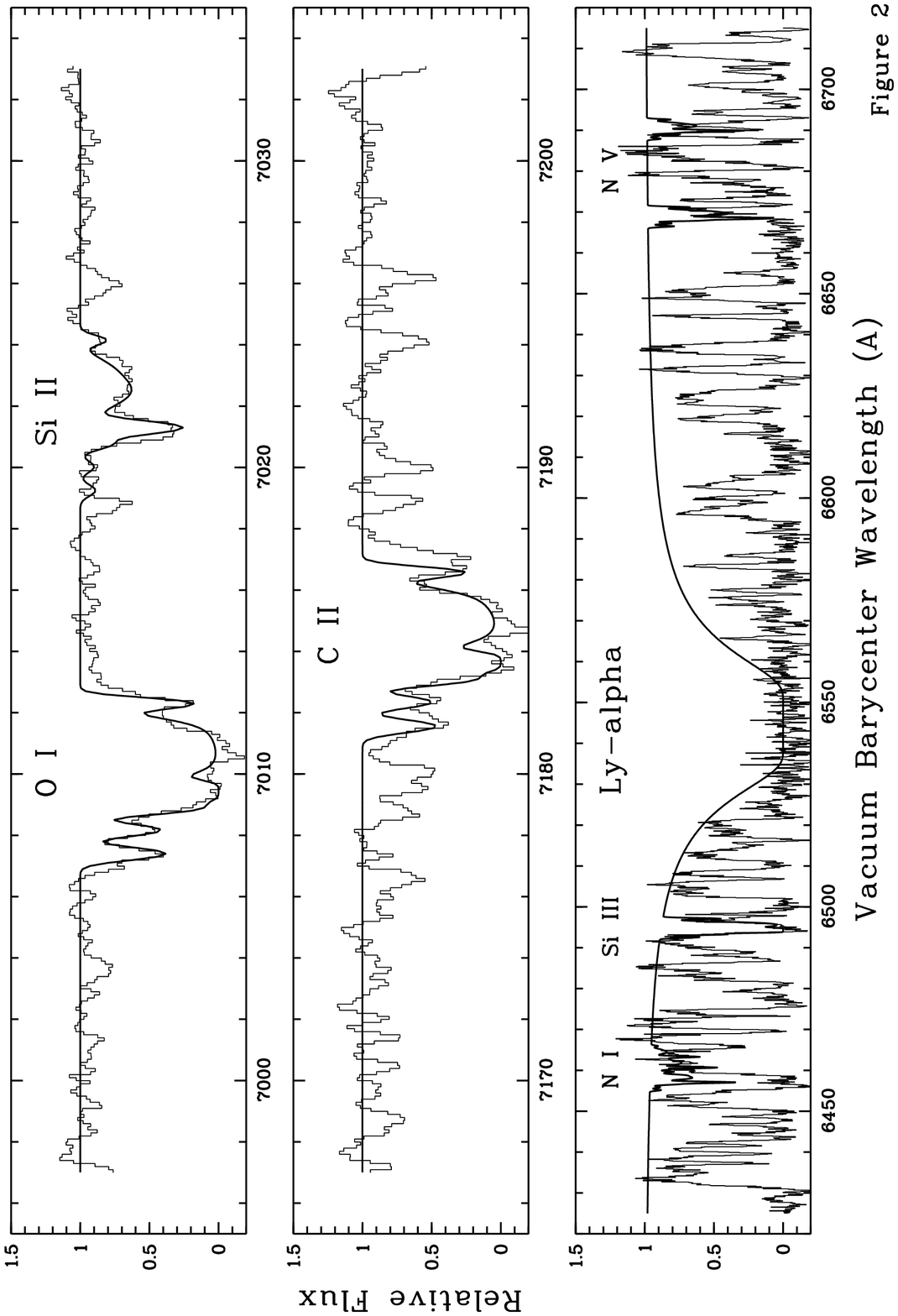}
\end{figure}

\begin{figure}
\vspace*{14cm}
\includegraphics{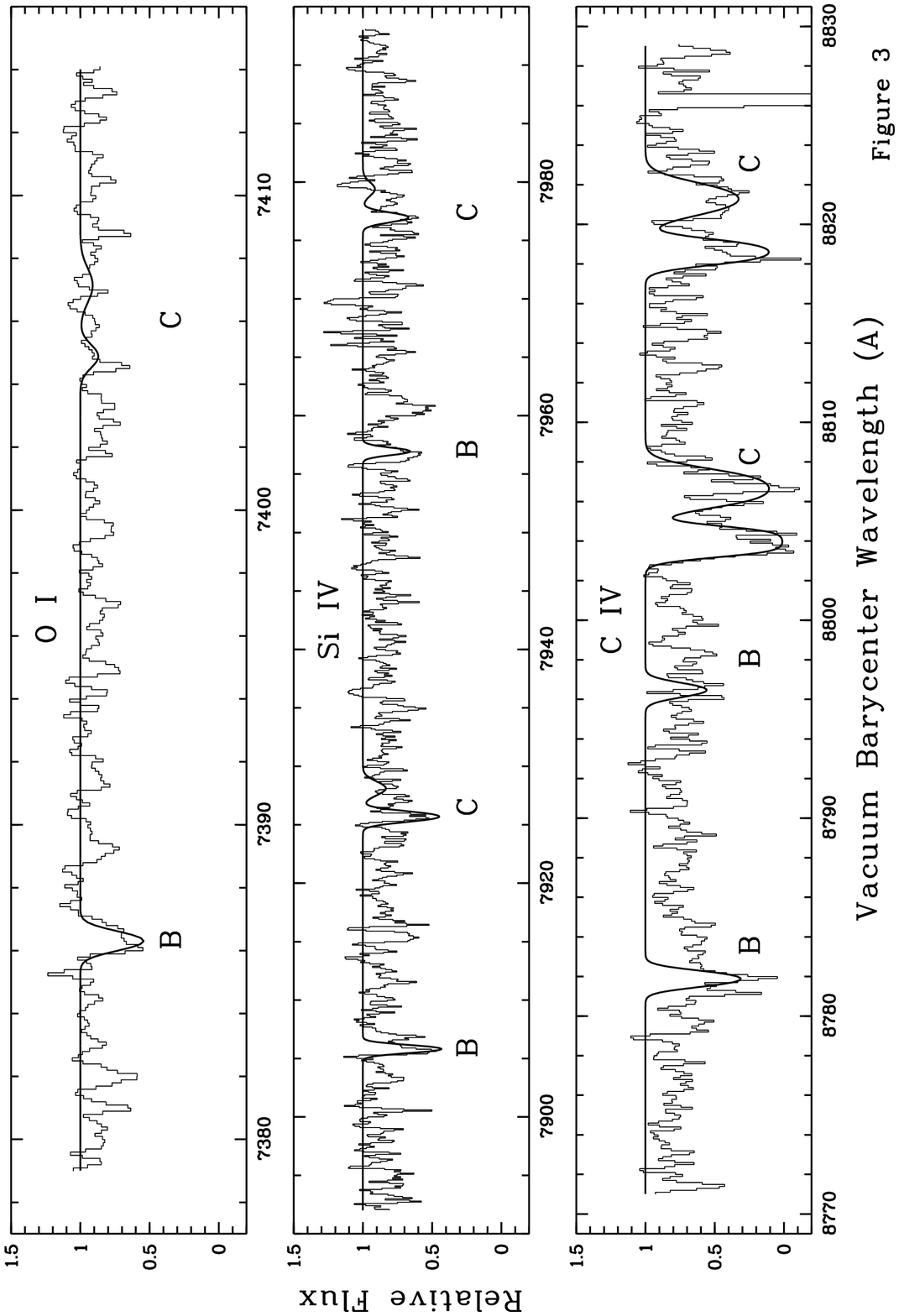}
\end{figure}

\begin{figure}
\vspace*{14cm}
\includegraphics{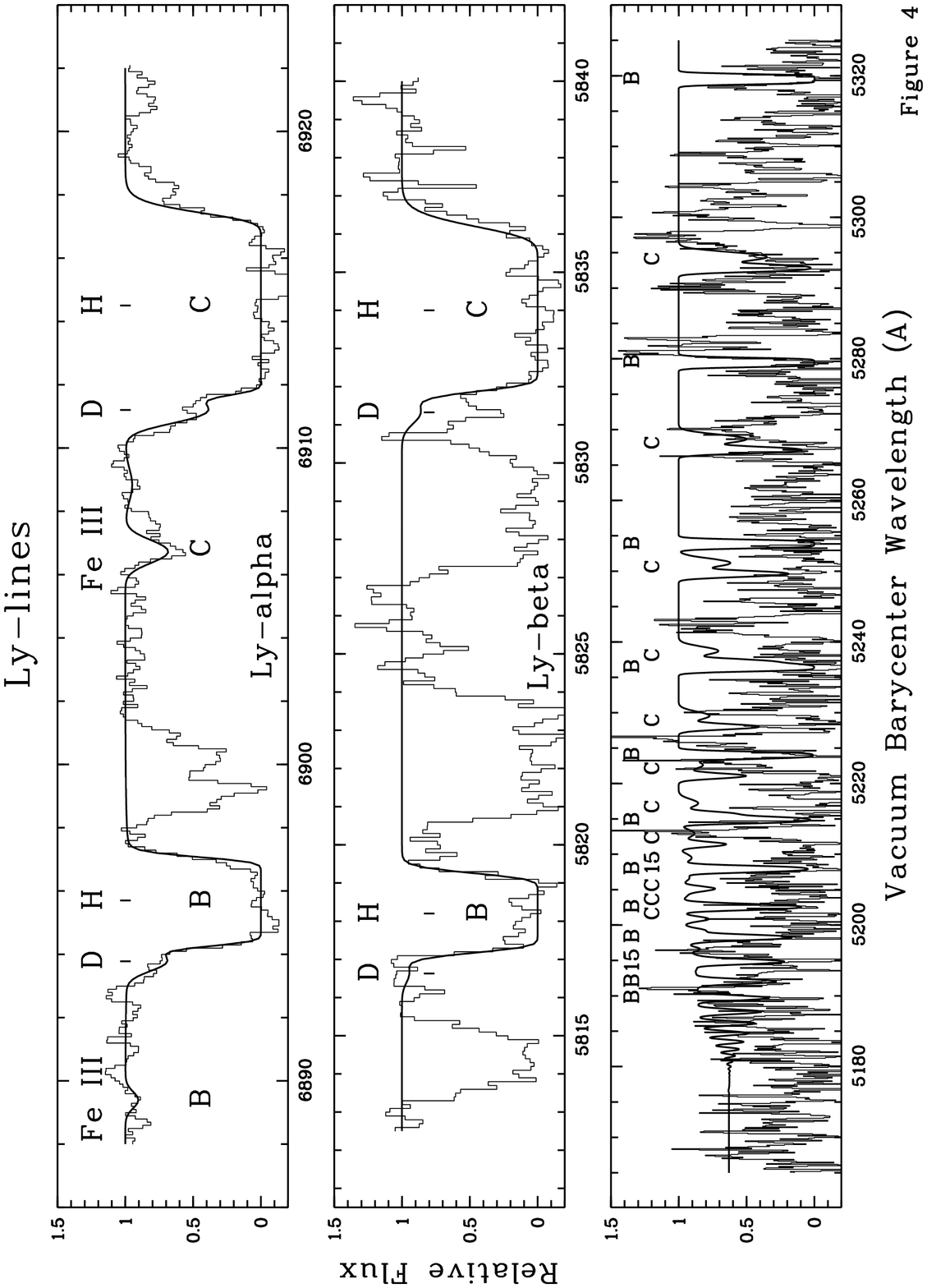}
\end{figure}

\begin{figure}
\vspace*{14cm}
\includegraphics{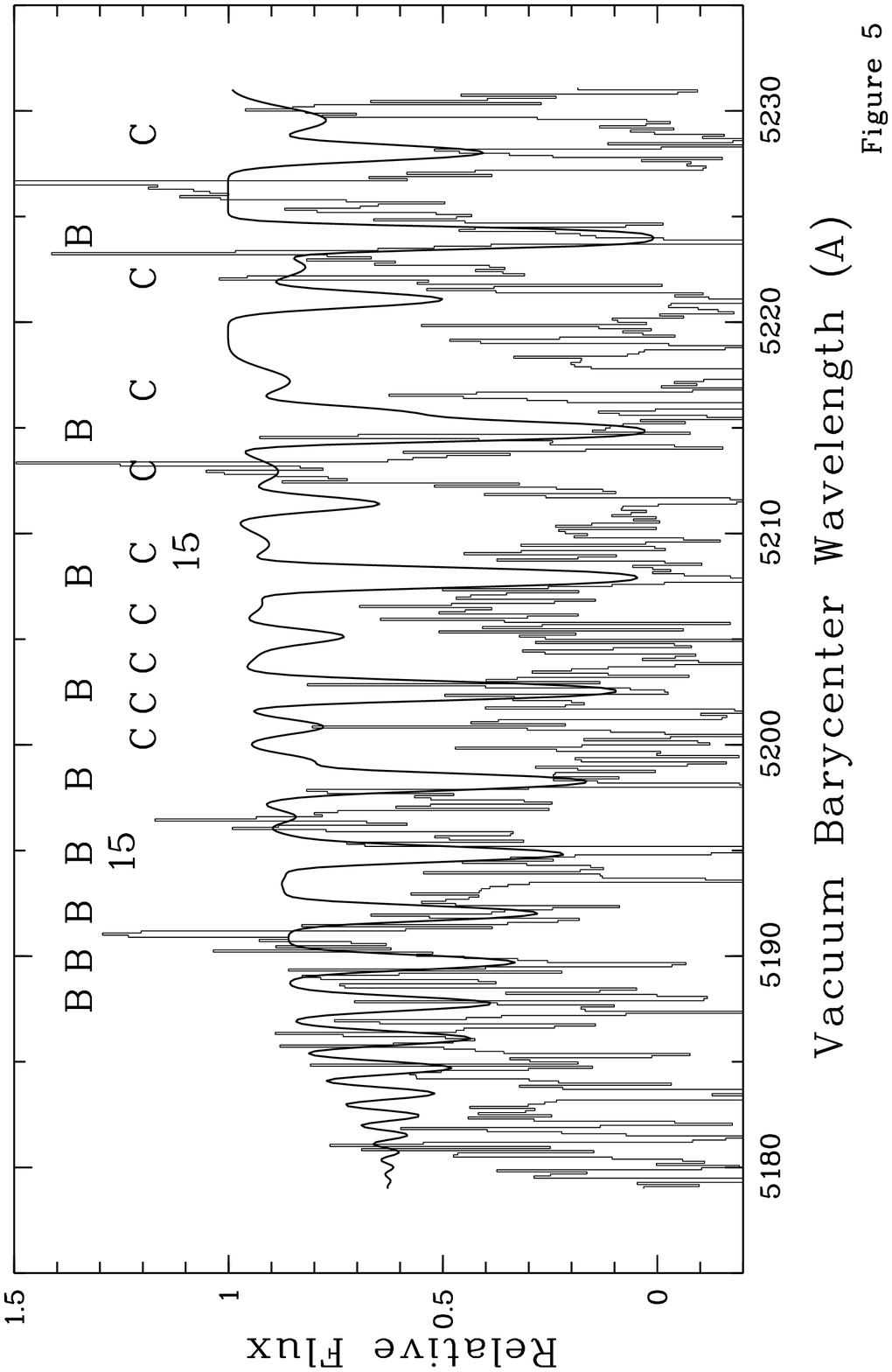}
\end{figure}

\end{document}